\documentclass[11pt,twoside,a4paper]{article}

\usepackage[top=20mm, bottom=25mm, left=25mm, right=25mm]{geometry}
\usepackage{graphicx}
\usepackage{titling}

\date{}

\newcommand{\affil}{}
\newcommand{\affiliation}[1]{
  \renewcommand{\affil}{#1}
}

\makeatletter
\renewcommand{\maketitle}{
  \begin{center}
    {\Large \@title}\\[3mm]
    {\@author}\\[3mm]
    {\affil}
  \end{center}
}
\makeatother

\newcommand{\chapter}[1]{
  \newpage
  \vspace*{80mm}
  \begin{center}
    \bf\huge 
    Chapter:\\#1
  \end{center}
}

\begin{document}

  \title{High-energy resummation effects in Mueller-Navelet jets production at the LHC}
\author{Bertrand Duclou\'e$^1$, Lech Szymanowski$^2$, Samuel Wallon$^{3,4}$}
\affiliation{$^1$Department of Physics, University of Jyv\"askyl\"a, P.O. Box 35, 40014 University of Jyv\"askyl\"a, Finland \\
             $^2$National Centre for Nuclear Research (NCBJ), Warsaw, Poland \\
             $^3$LPT, Universit{\'e} Paris-Sud, CNRS, 91405, Orsay, France \\
             $^4$UPMC Univ. Paris 06, Facult\'e de Physique, 4 place Jussieu, 75252 Paris Cedex 05, France}

\maketitle

The high energy limit of QCD, described by the Balitsky-Fadin-Kuraev-Lipatov (BFKL)
approach~\cite{Fadin:1975cb,Kuraev:1976ge,Kuraev:1977fs,Balitsky:1978ic}, has been the subject of many studies since four decades. One of the most promising processes to study these dynamics at hadron colliders was proposed by Mueller and Navelet~\cite{Mueller:1986ey}.
Consider two jets  separated by a large rapidity interval, i.e. each of them almost fly in the direction of the hadron ``close'' to it, and with similar transverse
momenta. Besides the cross-section, much information on the QCD high-energy resummation effects can be obtained when studying the azimuthal correlations of the two jets~\cite{DelDuca:1993mn,Stirling:1994zs}.
In a pure leading order (LO) collinear treatment, the two jets should be emitted back to back since there is no phase space for (untagged) emission between them. This simple picture is of course corrected by the inclusion of radiative corrections.
On the other hand, in the high-energy limit, the multiple emission of semi-hard gluons between these two jets is expected to modify dramatically this picture, leading to enhanced cross-sections and strong decorrelation effects. However, it is known that passing from a leading logarithmic (LL) to a next-to-leading-logarithmic (NLL) treatment in the BFKL framework can modify significantly this picture. 

Technically,
such a BFKL treatment involves two building blocks:
 the jet 
vertex, which describes the transition from an incoming parton to a jet,
and the Green's function, which describes the pomeron exchange between the vertices. A complete NLL BFKL analysis of Mueller-Navelet jets (for more details, see refs.~\cite{Ducloue:2013hia,Ducloue:2013bva}), including the NLL corrections both to the Green's function~\cite{Fadin:1998py,Ciafaloni:1998gs} and to
the jet vertex~\cite{Ciafaloni:1998kx,Ciafaloni:1998hu,Bartels:2001ge,Bartels:2002yj,Caporale:2011cc,Hentschinski:2011tz,Chachamis:2012cc}, showed that the NLL corrections to the jet vertex have a very large effect, of the same order of magnitude as the NLL corrections to the Green's function~\cite{Vera:2007kn,Marquet:2007xx}, leading to a lower cross-section and a much larger azimuthal correlation~\cite{Colferai:2010wu}. However, these results are very dependent on the choice
of the scales, especially the renormalization scale $\mu_R$ and the
factorization scale $\mu_F$, in particular in the case of realistic kinematical cuts for LHC experiments~\cite{Ducloue:2013hia}. This dependency can be reduced by using the Brodsky-Lepage-Mackenzie (BLM) prescription~\cite{Brodsky:1982gc}, adapted to the resummed perturbation theory \`a la BFKL~\cite{Brodsky:1998kn,Brodsky:2002ka}, to fix the renormalization scale. Such a full NLL BFKL analysis supplemented by the BLM scale fixing procedure has been performed~\cite{Ducloue:2013bva},
leading to a very satisfactory description of
the most recent LHC data extracted by the CMS collaboration
for the azimuthal correlations of Mueller-Navelet jets at a center-of-mass energy $\sqrt{s}=7$ TeV~\cite{CMS-PAS-FSQ-12-002}.
Recently, it was shown that energy-momentum non conservation should not affect this NLL BFKL analysis significantly~\cite{Ducloue:2014koa}.

In the very large center-of-mass energy of the LHC, the contributions
of partons with very small $x$ fraction of the protons are dominating, and their distributions are therefore enhanced. This is a typical situation where multipartonic scattering contributions could be sizable, in particular when dealing with small transverse momenta of the tagged jets (which hopefully will become accessible at CMS and ATLAS) for which the usual twist suppression is not valid~\cite{Diehl:2011yj}.
In such a situation, the cross-section gets an additional contribution where each jet is emitted by a separate BFKL chain.
The exchange of two BFKL ladders is presumably the dominant contribution for asymptotical values of $s$, and it would be of great interest to determine at which point it starts to be of the same order of magnitude as the exchange of a single ladder.
Thus, the study of this process with transverse momenta of the tagged jets as low as possible would be a great opportunity towards investigating the effect of 
multipartonic contributions at small $x$.

  \setcounter{section}{0}

\end{document}